\documentstyle[12pt]{article}
\begin{document}
\begin{titlepage}
\rightline{IMSc 92/27}
%\rightline{\today}
\vskip1cm
\centerline{\large\bf O(3) Non-linear $\sigma$ model with Hopf term and}
\centerline{\large\bf  Higher spin theories}
\vskip1cm
\centerline{\bf T.R.Govindarajan$^1$, R.Shankar$^2$}
\centerline{The Institute of Mathematical Sciences, C.I.T. Campus,
Madras-600113, INDIA.}
\centerline{\bf N.Shaji}
\centerline{Centre for Development of Imaging Technology,
Trivandrum-695027, INDIA.}
\centerline{\bf M.Sivakumar$^3$}
\centerline{School of Physics, University of Hyderabad, Hyderabad-500134,
INDIA.}
\vskip1cm
\centerline{Abstract}

Following our earlier work we argue in detail for the equivalence of the
nonlinear $\sigma$ model with Hopf term at~$\theta=\pi/2s$ ~and an
interacting spin-s theory. We give an ansatz for spin-s
operators in the $\sigma$ model and show the equivalence of the correlation
functions.We also show the
relation between topological and Noether currents. We obtain
the Lorentz and discrete transformation properties of the
spin-s operator from the fields of the $\sigma$ model.
We also explore the connection of this model with Quantum Hall Fluids.
\vfill
\hrule
\noindent{\em
email:~$^1$trg,$^2$shankar@imsc.ernet.in;~$^3$ms-sp@uohyd.ernet.in}
\vskip1cm
\end{titlepage}
\noindent{\bf 1.Introduction:}
\vskip.75cm
Interest in the $2+1$ dimensional $O(3)$ nonlinear $\sigma$ model (NLSM) with
the Hopf
term was roused when Wilzcek and Zee\cite {WZ} pointed out that the Hopf
term\cite{H} causes the solitons
to acquire fractional spin and statistics. The connection of the NLSM  to the
long wavelength fluctuations of antiferromagnets  had been established by
Haldane
\cite{HA}.Interesting possibilities of the model without the Hopf term
has been discussed by Wen and Zee \cite{WeZ}. The Hopf term has not been
derived from a microscopic  spin model so far\cite{REF}. But it is
expected on physical grounds that in some classes frustrated antiferromagnets
the Hopf-term
may actually be generated\cite{FRA}.

The topological features of this model gives the solitons the fractional
statistics and spin.
This model was analysed in a canonical framework by Bowick,Karabali and
Wijayavardhana\cite{BO}
wherein they pointed out that addition of Hopf term adds a term to
the angular momentum and changes the statistics of the solitons.Later the same
model was analysed by Dzyaloshinsky,
Polyakov and Wiegmann\cite{DPW} who showed that in the long wavelength limit
the effect of
Hopf term is to add a Chern-Simon's term to the model. Further topological
features of the
model with torus and arbitrary genus Riemann surface boundary conditions gives
the
possibilities of fractional statistics only by using multi component
wavefunctions\cite{OUR1,OUR2}.
Following $CP_1$ formalism  Polyakov\cite{PO} then showed relativistic
particles
interacting with an abelian Chern Simons gauge fields (with $\theta~=~\pi$)
are Dirac fermions in the long wavelength limit. This can easily be generalized
to
higher integer and half odd integer spins as well\cite{AL}. Some of us then
showed that the
above result holds independent of long wavelength  approximation\cite{SSS}  and
also in the
presence of self interactions\cite{SS}

In an earliar paper\cite{GSSS} we argued that this model for $\theta~=~
{\pi\over 2s},~ ~s={1\over 2},1,{3\over 2},\cdots$ is equivalent
to an interacting spin s theory. In particular when $s = {1\over 2}$ the theory
is
equivalent to Dirac fermions with four fermi interactions. In this paper
we present this argument in full detail and give further consequences.

In Sec.2 we present a brief review of the topological features of the model
and the origin of the fractional statistics and spin of the solitons. In Sec.3
We
present a local $CP_1$ model involving two gauge fields  which is exactly
equivalent
to O(3) $\sigma$ model with Hopf term. We present the route to higher spin
formulation by formally integrating the gauge fields and $CP_1$ fields and show
that
the partition function is the same as obtained by integrating the higher spin
fields in an
interacting higher spin theory in Sec.4. We also establish in Sec.5 the
equivalence by relating the
correlation functions and in Sec.6 bring  out  the relationship between
topological current
and the particle (Noether) current of the higher spin theory. Finally we derive
the
transformation properties of higher spin fields from the known transformation
properties
of the $CP_1$ fields and gauge fields in Sec.7. In the final Sec.8 we conclude
 with discussions of our work.Appendix A contains path integral
representation for spin-s pariticles while in Appendix B
we discuss the transformation properties of the spin-s fields.
\vskip.75cm
\noindent {\bf 2.Topological features of the model:}
\vskip.75cm
The model we consider is described by a three dimensional unit vector {\bf n}
field
with $n^an^a = 1$ with $a=1,2,3$ in 2+1 dimensions. It is described by the
Euclidean
action
\begin{equation}
S = g^2\int \partial_\mu n^a \partial_\mu n^a+ i\theta H(n^a)
\end{equation}
Here g is the coupling constant and the second term in the action is the Hopf
invariant.
At any instant of time for finite energy solutions the boundary conditions on
the
{\bf n} are such that
$$
n\longrightarrow (0,0,1)
$$
Because of this the 2d plane becomes the compact $S^2$. Hence the configuration
space for this field theory is
$$
Q\equiv Set~of~maps~from~S^2~\longrightarrow~S^2
$$
This space is not path connected as can be seen from the following:
\begin{equation}
\pi_0(Q) \equiv \pi_2(S^2) = {\cal Z}
\end{equation}
Hence there exists solitons in this model and soliton number is conserved.
There
exists the topological current given by
\begin{equation}
j_\mu^{top} = {1\over
8\pi}\epsilon_{\mu\nu\lambda}\epsilon_{abc}n_a\partial_\nu n_b \partial_\lambda
n_c
\end{equation}
which is conserved. $j_0^{top}$ is the soliton density.

If we consider $\pi_1(Q)$ then it is easy to see that
\begin{equation}
\pi_1(Q) \equiv \pi_3(S^2) = {\cal Z}
\end{equation}
Hence the quantization is ambiguous \cite{TRG} upto a phase $e^{i\theta}$. The
effect of such an ambiguity
can be incorporated in the action by adding the second term in (1), the Hopf
term.

{}From the fact that $j_\mu^{top}$ is conserved one can write down, atleast in
zero
soliton sector,
\begin{equation}
j_\mu^{top} = \epsilon_{\mu\nu\lambda}\partial_\nu A_\lambda
\end{equation}
Given such a vector potential $A_\mu$, the Hopf term is given by
\begin{equation}
H (n) = \int j_\mu^{top} A_\mu d^3x
\end{equation}
Note that Hopf term (6) is non local since $A_\mu$ defined in (5) has to be
non local in the {\bf n} fields.

For any configuration which evolves in time the Hopf invariant arises in the
following way. A typical soliton configuration in the 2d plane be represented
by a disc of radius R centered around some point, say origin. It is given by
$$
n\cdot \tau = \tau^3 cos f(r) + {\hat x\cdot \tau sin f(r)}
$$
with
$$
f(0) = \pi;~~~~ f (R) = 0.
$$
Such a soliton configuration in time generates a tube. In the zero soliton
sector, a soliton and an anti soliton can be created and annihilated as shown
in Fig.1.
\setlength{\unitlength}{1cm}
$$
\begin{picture}(5,4.5)
\put(0,0){\line(1,0){4}}
\put(0,4){\line(1,0){4}}
\put(.5,.5){\line(1,0){3}}
\put(.5,3.5){\line(1,0){3}}
\put(0,0){\line(0,1){4}}
\put(.5,.5){\line(0,1){3}}
\put(3.5,.5){\line(0,1){3}}
\put(4,0){\line(0,1){4}}
\put(3.75,1){\vector(0,1){2}}
\put(.25,3){\vector(0,-1){2}}
\put(2.0,-.6){Fig.1}
\end{picture}
$$
\vskip.5cm
%\centerline{\em Fig.1}
This configuration has Hopf invariant 0. But if we rotate the soliton by $2\pi$
before annihilating we get Hopf invariant 1 correspoding to the configuration
in Fig.2.
$$
\begin{picture}(5,4.5)
\put(0,0){\line(1,0){4}}
\put(0,4){\line(1,0){4}}
\put(.5,.5){\line(1,0){3}}
\put(.5,3.5){\line(1,0){3}}
\put(0,0){\line(0,1){4}}
\put(.5,.5){\line(0,1){3}}
\put(4,0){\line(0,1){.5}}
\put(4,3.5){\line(0,1){.5}}
\multiput(3.5,.5)(0,1.5){2}{\line(1,3){.5}}
\multiput(3.8,1.1)(0,1.5){2}{\line(1,-3){.2}}
\multiput(3.5,3.5)(0,-1.5){2}{\line(1,-3){.2}}
\put(.25,3){\vector(0,-1){2}}
\put(4.2,1){\vector(0,1){2}}
\put(2.0,-.6){Fig.2.}
\end{picture}
$$
\vskip.5cm
%\centerline{\em Fig.2.}
This accounts for the spin of the soliton. The amplititude for this
process has a phase ${\theta}$ compared to that in Fig.1.If we take soliton
anti soliton pairs
created  at two different locations exchange and further
annihilation of the solitons give the configuration Fig.3.
\setlength{\unitlength}{.8cm}
$$
\begin{picture}(12,5)
\put(0,0){\line(1,0){4}}
\put(0,4.5){\line(1,0){4}}
\put(.5,.5){\line(1,0){3.5}}
\put(.5,4){\line(1,0){3.5}}
\put(0,0){\line(0,1){4.5}}
\put(.5,.5){\line(0,1){3.5}}
\put(8,0){\line(1,0){4}}
\put(8,4.5){\line(1,0){4}}
\put(8,.5){\line(1,0){3.5}}
\put(8,4){\line(1,0){3.5}}
\put(12,0){\line(0,1){4.5}}
\put(11.5,.5){\line(0,1){3.5}}
\multiput(4,0)(0,.5){2}{\line(1,1){4}}
\multiput(4,4.5)(0,-.5){2}{\line(1,-1){1.7}}
\multiput(8,0)(0,.5){2}{\line(-1,1){1.7}}
\put(.25,3.5){\vector(0,-1){2.5}}
\put(11.75,3.5){\vector(0,-1){2.5}}
\put(4.5,.75){\vector(1,1){3}}
\put(5.5,-.6){Fig.3.}
\end{picture}
$$
\vskip.5cm
%\centerline{\em Fig.3.}
This can be isotopically mapped onto the previous one and thereby accounting
for the statistics  of the solitons. Hence the addition of the Hopf term leads
to
the change of spin and statistics of the solitons of the model. If we
had considered the same process with solitons of soliton number $n_s$
the Hopf invariant for configurations in Fig.2 and Fig.3. is $n_s^2$
and hence the spin and statistics of soliton number $n_s$ object
is
\begin{equation}
Spin~~=~~n_s^2 {\theta \over 2\pi},~~~~~Statistics~=n_s^2\theta \label{nsol}
\end{equation}
\vskip.75cm
\noindent{\bf 3.$CP_1$ formalism}
\vskip.75cm
We now present the $CP_1$ formalism in which {\bf n} fields are represented
through $CP_1$
fields. This has the advantage that the Hopf term is local in $CP_1$ formalism.
We use the fact
$$
S^2 ~~ = ~~ {SU(2)\over U(1)}
$$
and reexpress {\bf n} fields in terms of $SU(2)$ matrices U through the
definition
$$
U^\dagger \tau^3 U ~~ = ~~ n^a \tau^a
$$
Here $\tau$ are Pauli matrices. For convenience we parametrize U through two
complex fields $z_1,z_2$.
$$
U~~=~~ {1\over 2g} \pmatrix {z_1 & z_2 \cr -z_2^* & z_1^*}
$$
U being an unitary matrix we get,
$$
\sum_{\sigma=1}^2 |z_\sigma|^2 ~~=~~4g^2
$$
We also define
$$
L_\mu~~=~~i\partial_\mu U U^\dagger,~~~ L_\mu^a~=~{1\over 2}tr~ \tau^a L_\mu
$$
In terms of $L_\mu$, the elements of $SU(2)$ algebra, we get
$$
g^2 \partial_\mu n^a \partial_\mu n^a ~=~4g^2(L_\mu^1L_\mu^1~+~L_\mu^2L_\mu^2)
{}~=~\sum|\partial_\mu z_\sigma|^2~-~4g^2L_\mu^3L_\mu^3.
$$
{}From the definition of $j_\mu^{top}$ we find that
\begin{equation}
j_\mu^{top}~=~{1\over 2\pi}\epsilon_{\mu\nu\lambda}\partial_\nu L_\lambda^3.
\end{equation}
Hence the Hopf term becomes
\begin{equation}
H(n)~~=~~ -{1\over 4\pi^2}\epsilon_{\mu\nu\lambda}L_\mu^3\partial_\nu
L_\lambda^3.
\end{equation}
The action of the non-linear sigma model can be rewritten in this $CP_1$
formalism
as,
\begin{equation}
S_{CP_1}~~=~~\int |\partial_\mu
z_\sigma|^2~-~4g^2L_\mu^3L_\mu^3~-~{i\theta\over 4\pi^2}
\epsilon_{\mu\nu\lambda}L_\mu^3\partial_\nu L_\lambda^3.
\label{a}\end{equation}
Next we introduce auxiliary fields $a_\mu$ and $b_\mu$ as follows:
$$
exp(4g^2\int L_\mu^3L_\mu^3)~=~\int_{b_\mu}~exp~(4g^2(2b_\mu L_\mu^3~-~b_\mu
b_\mu))
$$
\begin{equation}
exp~({i\theta\over 4\pi^2}\int \epsilon_{\mu\nu\lambda}L_\mu^3\partial_\nu
L_\lambda^3)
{}~=~\int_{a_\mu} exp~({i\theta\over
4\pi^2}(2\epsilon_{\mu\nu\lambda}a_\mu\partial_\nu
L_\lambda^3~-~\epsilon_{\mu\nu\lambda}a_\mu\partial_\nu a_\lambda))
\label{b}\end{equation}
Notice that left hand side of the second equation is periodic in $\theta$ due
to the
integer valued Hopf invariant. But the periodicity of $\theta$ on the right
hand side
is not manifest. But if we make the shift
$$
a_\mu \longrightarrow a_\mu'~=~ ca_\mu~+~(1-c)L_\mu^3
$$
where
$c~=~(1+{2n\pi\over\theta})^{1\over 2}$ then this has the effect of replacing
$\theta$ by $\theta~+~2n\pi$ in the action for the $a_\mu$ field and one
recovers
periodicity.

We also have the constraint
$$
z^\dagger z~=~4g^2
$$
This constraint is introduced through the lagrange multiplier $\lambda$ in the
action
as $\lambda(z^\dagger z~-~4g^2)^2$. Using all these the $CP_1$ action is given
by,
\begin{equation}
S_z~~=~~\int z^\dagger Gz ~+~S_\eta~+~S_{GF}
\label{z}\end{equation}
where
\begin{eqnarray*}
G~&=& -D_\mu D_\mu -2i\eta \sqrt {\lambda}\\
D_\mu~&=&i\partial_\mu~-~A_\mu\\
A_\mu~&=&b_\mu~+~i\alpha \epsilon_{\mu\nu\lambda}\partial_\nu a_\lambda\\
\alpha~&=&{\theta\over 16\pi^2g^2}
\end{eqnarray*}
and
\begin{equation}
S_{GF}~=~4g^2\int (-2i\alpha \epsilon_{\mu\nu\lambda}b_\mu\partial_\nu
a_\lambda~+~
\alpha^2(\epsilon_{\mu\nu\lambda}\partial_\nu
a_\lambda)^2~+~i\alpha\epsilon_{\mu\nu\lambda}
a_\mu\partial_\nu a_\lambda)
\end{equation}
$$
S_\eta~=~\int (\eta^2~-~8i\sqrt{\lambda}\eta g^2)
$$
This theory defined by this action is formally exactly equivalent to NLSM in
the
limit $\lambda \longrightarrow \infty$ and we shall work with this action
which we refer to as z-theory from now on.
\vskip.75cm
\noindent{\bf 4.Equivalence of partition functions:}
\vskip.75cm
First we analyze the partition function of the model.
The $z$ fields, occurring quadratically in the partition function,
can be
integrated out and we have,
\begin{equation}
{\cal Z}~=~\int
e^{-S}~=~\int_{a_\mu,b_\mu,\eta}~exp\{-2ln~det~G~-S_{GF}~-~S_\eta\} .
\end{equation}
We can use the heat kernel representation of the logarithm of the determinant.
$$
-2~ln~det~G~=~2\int_{1\over \Lambda^2}^{\infty}{d\beta\over \beta}~e^{-\beta
G}~~~~~~~~~~~~~~~~~
$$
\begin{equation}
{}~~~~~~~~~~~~~~~~=2\int_{1\over \Lambda^2}^{\infty}~{d\beta\over
\beta}\int_{x(\tau)}~e^{-
\int_0^\beta d\tau ({1\over 4}(\partial_\tau x^\mu)^2~+~V(x)~-~i\oint A_\mu
dx^\mu }.
\label{pi}
\end{equation}
where we have defined $V(x)~\equiv ~2i\sqrt{\lambda}\eta $ and $ \Lambda$ is
the ultraviolet
cut off.
We see that the dependence of the gauge field in the determinant
operator, is
of the form of Wilson loop, which is of course expected, due to the gauge
invariant nature of the `Hamiltonian' $G$. Expanding the exp$-2 ln~G$
in
power series, the partition can be written as
$$
{\cal Z}\ = \ \sum^\infty_{n=0} \ \frac{{\cal Z}_n[\eta]}{n!}e^{- S_\eta}
$$
where
$$
{\cal Z}_n [\eta] \ = \ \prod^n_{i=1} 2^n \int_{\Lambda^{-2}}^\infty
\frac{d\beta_i}{\beta_i}
\int Dx_i e^{i \sum^n_{i=1} \int^{\beta_i}_0 d\tau \left[ \frac{1}{4}
\left(
\partial_\tau x^\mu_i \right)^2 + V(x_i) \right]}
$$
\begin{equation}
{}~~~~~\times \int DaDb e^{- i \oint_{C} Adx_i - S_{GF}}
\label{gf}\end{equation}

This describes a grand canonical ensemble of particles, with ${\cal Z}_n$
interpreted as trajectories of $z$ - particles. First we integrate over
the
gauge fields.
The averaging over gauge fields of an arbitrary product of Wilson loops
can be done as follows. Let $C_i~i~=1,2,...n$ represent n distinct loops and
$C\equiv
\bigcup C_i$ corresponding to propagation of $n$ pairs of particles
antiparticles
created and destroyed after travelling time $\beta_i$,The $b_\mu$
integrals can now done to obtain
$$
\int_{b_\mu} e^{-i \oint A_\mu dx^\mu ~-~S_{GF}} \ = \ S(j_\mu^c
{}~-~\frac{\theta}{2\pi^2} \epsilon_{\mu \nu \lambda} \partial_\nu a_\lambda
(x))
\hskip2cm $$
\begin{equation}
\qquad \times exp \int_x \left[2 d_{\mu\nu}a_\mu j^c_\nu - 4g^2
(d_{\mu\nu}a_\nu)^2 - 4ig^2a_\mu d_{\mu\nu}a_\nu\right]
\label{jflux}
\end{equation}
where
$$
j_\mu^c \ = \ \sum^n_{i=1} \int \partial_\tau x_\mu^{c_i} \delta^3 (x -
x_\mu^{c_i}(\tau))
$$
and $d_{\mu\nu}~=~\epsilon_{\mu\lambda\nu}\partial_\lambda$.
Now the $a_\mu$ integrals can be done easily. However, note that, at $\theta
= 0,
j^C_\mu =0$. This means that the trajectory of the particle along a given
curve, is accompanied by that of anti-particle, so that the total current
is
zero. Thus particles and antiparticles are confined and single $z$ particle
cannot propagate. The presence of the $`\theta'$ term leads to the
possibility
of deconfinement of $z$-particles, similar to what happens in Chern
Simons gauge
theories coupled to matter-fields as discussed in \cite{DPW}.
When $a_\mu$ integrals are also done, we obtain
\begin{equation}
exp~(\int -~{i\pi^2\over\theta}j\cdot d^{-1}\cdot j
{}~+~{1\over 16g^2}~j\cdot j)
\label{jj}\end{equation}
where $j\cdot d^{-1}\cdot j \equiv
{}~j_\mu(\epsilon_{\mu\nu\lambda}\partial_\nu)^{-1}j_\lambda.
$ Hence we get,
$$
\langle\langle e^{-\oint A_\mu dx^\mu}\rangle\rangle~=~exp(-{i\pi^2\over
\theta}{1\over 4\pi}
\oint dx^\mu\oint dy^\nu\epsilon_{\mu\nu\lambda}{(x-y)_\lambda\over
|x-y|^3}~+~{1\over 16g^2}
\int_x j_\mu^Cj_\mu^C)
$$
The first term in this is the well known integral and is given by:
\begin{equation}
exp({i\pi^2\over\theta}(\sum_1^n W(C_i)~+~ \sum_{i\neq j}2n_{ij}))
\end{equation}
where
$W(C_i)$ is the writhe of the curve $C_i$ and $n_{ij}$ is the linking number of
the
curves $C_i$ and $C_j$. For $\theta~=~{\pi\over 2s} $ the linking number term
does not
contribute. The writhe $W(C_i)$ has the expression,
\begin{equation}
W(C_i)~=~{1\over 2\pi}\Omega (C_i)~+~2k+1
\end{equation}
where $\Omega (C_i)$ is the solid angle subtended  on a 2-sphere traced
out by the unit tangent vector to $C_i$ and $2k+1$ is an odd integer. Hence
$$
e^{-2\pi~is~W(C_i)}~=~(-1)^{2s}~e^{-is~\Omega(C_i)}
$$
Again linearizing the second term of Eq.(\ref{jj}) and combining the result of
integrating z,$a_\mu$ and $b_\mu$ we can write the partition function as
\begin{equation}
{\cal Z}~=~\int_{\eta,v_\mu}~e^{-S_\eta~-~\int_xv_\mu v_\mu~-~2\int
{d\beta\over\beta}\int_x
e^{-\int d\tau{\dot x^2_\mu\over 4}~+~V(x)~+~(-1)^{2s}is~\Omega}}
\end{equation}
%To do this path integral we introduce the velocity variables,
%$$
%u_\mu~=~\partial_\tau x_\mu(\tau)
%$$
%and lagrange multiplier $k_\mu$ to impose the constraint.

The addition of the Polyakov spin factor, $e^{is \Omega}$ to the free
Bosonic path integral has been studied \cite{shaji} well in recent times, and
has been shown
to describe spinning particles, with spin $s$. We have here, a similar set up,
except for the inclusion of background scalar $(V(x))$ and vector $(v_\mu)$
fields. In the Appendix-A, we extend this proof, of showing its equivalence to
spin-$s$ particle, in the presence of external fields. Using the results of the
appendix, we have
$$
{\cal Z}_n[\eta] = 2^n \int_{v_\mu} e^{\int u^2_\mu} \prod^n_{i=1} (-)^{2s}
\int^\infty_{\Lambda^{-2}} \frac{dL_i}{L_i} Tr e^{- L_i D^{(s)}}
$$
where
$$
L \ = \ \frac{\beta}{\kappa} \qquad \kappa = \frac{\sqrt{\pi}}{4\Lambda}
$$
$$
D^{(s)} \ = \ sgn(\theta) (i\partial_\mu + i \frac{1}{2g} v_\mu)
\frac{T^\mu}{s} + M_s +
\kappa V(x)
$$
$$
M_s \ = \ \Lambda^2 \kappa~ln(2s +1)
$$
The mass term $(M_s)$, of $O(\Lambda)$, arises from $\epsilon$ the  distance
between adjacent $\tau$-slices, being equal to $\frac{1}{\Lambda^2}$.

Doing the $L$ integral, we get
$$
{\cal Z}_n \ = \ \int_{v_\mu} e^{-\int_\lambda v_\mu v_\mu} 2^n (-)^{2s}~ln
{}~det(D^s)^n
$$
We can now substitute in Eq.(\ref{gf}) and sum the series to get
$$
{\cal Z} \ = \ \int_{v_\mu,\eta} e^{- \int v_\mu v_\mu - \int \eta^2}
(-)^{2s+1} det
(D^s)^2
$$
Depending on $2s+1$ being odd or even, this determinant can be
written as functional integral over two-component complex $c$-number fields or
Grassmanian fields $\bar{\psi}_{m\sigma}, \psi_{m\sigma} (m = -s\cdots s,
\sigma = 1,2)$ respectively. The two-component isospin index $\sigma$, comes
from the two component nature of the original $z$ fields. The $v_\mu$ and
$\eta$ integrals can also be performed and we finally obtain for the partition
function,
\begin{equation}
{\cal Z} \ = \ \int_{\bar{\psi}_\sigma \psi_\sigma} exp - \{ \int_x \left[
\bar\psi
\left( \frac{T^\mu}{s}~i\partial_\mu + M \right) \psi~+~r_1 \left(\bar{\psi}
T^\mu \psi \right)^2 + r_2 \left( \bar{\psi} \psi \right)^2 \right]\}
\label{hs}\end{equation}
where
$$
M \ = \ sgn(\theta)(M_s - 8 g^2 \kappa \lambda)
$$
$$
r_1 \ = \ \frac{1}{16g^2}
$$
$$
r_2 \ = \ \lambda \kappa^2
$$
Here we have absorbed the factor $sgn(\theta)$ with the fields. Thus it
now appears with the mass term.
Thus the partition function of the $z$-theory is equal to that of the
above spin-$s$ theory.
\newpage
\def \d {{\rm d}} \def \D {{\cal D}}
\def \tr {{\rm tr}} \def \Tr {{\rm Tr}}
\noindent{\bf 5.Equivalence of correlation functions:}
\vskip.75cm
In this section we explicitly construct nonlocal fields in the NLSM and
prove that their two point correlation functions are exactly equal to the
two point correlation functions of the $\psi$ and $\bar\psi$ fields
in the spin-s theory. The proof can be easily generalised to n-point
functions following the methods in reference \cite {SSS}.

Our ansatz for spin-$s$ fields is as below,
\begin{eqnarray}
\chi_{\sigma u}(x)&=&\zeta[c]\exp\left(i\int_c^x L^3_\mu{\rm dx}_\mu\right)
\Delta(u_\mu-d_{\mu\lambda} L^3_\lambda(x))
z_\sigma(x)\\
\bar\chi_{\sigma u}(x)&=&\bar\zeta[c]
\exp\left(-i\int_c^x L^3_\mu{\rm dx}_\mu\right)
\Delta(u_\mu-d_{\mu\lambda} L^3_\lambda(x))
z^*_\sigma(x)
\label{anz}
\end{eqnarray}
Here the line integral is over a fixed curve $c$ from some  arbitrary
point $x_0$ to $x$ and $\zeta[c]$ is a normalisation factor, which
depends on the curve $c$ and will be chosen suitably. The $
\Delta(u_\mu-d_{\mu\lambda} L^3_\lambda)$ term
stands for an angular delta function, which can be got by integrating a
Dirac delta function over the radial variables. That is,
\begin{displaymath}
\Delta(u_\mu-d_{\mu\lambda} L^3_\lambda)
=\int_0^\infty {\rm u^2du}
\delta(u_\mu-d_{\mu\lambda}L^3_\lambda)~.
\nonumber
\end{displaymath}
Now if we use the standard integral representation for the delta
function, we can write,
\begin{eqnarray*}
\Delta(u_\mu-\epsilon_{\mu\nu\lambda}\partial_\nu L^3_\lambda)
&=&
\int\!\frac{\rm d^3k}{(2\pi)^3}\int_0^\infty{\rm u^2du}\nonumber\\
& &\times\exp\left(ik_\mu
(u_\mu-\epsilon_{\mu\nu\lambda}\partial_\nu L^3_\lambda)\right)~.
\nonumber
\end{eqnarray*}
The role of angular delta function in the ansatz is to relate the spin
index of the $\chi$ field to the direction of the topological current at
$x$. Now if we define
\begin{displaymath}
j^c_\mu=\int_c{\rm dx}_\mu\delta({\bf x-x(\tau))},\quad{\rm then}\quad
\int_c L^3_\mu{\rm dx}_\mu=\int\!{\rm d^3x}\,j^c_\mu L^3_\mu~.
\nonumber
\end{displaymath}
Let us also define $\tilde k_\mu=\delta(x-x_i)k_\mu$ so that we can write
$k_\mu\epsilon_{\mu\nu\lambda}L^3_\lambda$ as $\int\!{\rm d^3x}\,\tilde
k_{1\mu}\epsilon_{\mu\nu\lambda}L^3_\lambda$.
Now we can rewrite $\chi_{\sigma u}$ as
\begin{equation}
\chi_{\sigma u}=\zeta[c]\int\!\frac{\rm d^3k}{(2\pi)^3}\exp(ik_\mu u_\mu)
\exp(-ik_\mu\epsilon_{\mu\nu\lambda}\partial_\nu L^3_\lambda)
\exp(i\int\!{\rm d^3x}\,j^c_\mu L^3_\mu)~.
\label{chi}
\end{equation}

We will evaluate the two point correlation function with
this ansatz {(eq.\ref{chi})}and show that it is exactly equal to the two point
correlation function in the theory of interacting spin-$s$ fields.
\begin{equation}
\langle\bar\chi_{\sigma_1u_1}(x_1)\chi_{\sigma_2u_2}(x_2)\rangle
=
\int\!{\cal D}z{\cal D}z^*\exp(-S_{CP_1})
\bar\chi_{\sigma_1u_1}(x_1)\chi_{\sigma_2u_2}(x_2)~.
\label{chibarchi}
\end{equation}
Introducing the $a_\mu$ and $b_\mu$ fields as before, we get,
\begin{eqnarray}
\langle\bar\chi_{\sigma_1u_1}\chi_{\sigma_2u_2}\rangle
&=&
\frac{1}{\cal Z}\bar\zeta[c_{01}]\zeta[c_{02}]
\int\!\frac{\rm d^3k_1}{(2\pi)^3}
\int\!\frac{\rm d^3k_2}{(2\pi)^3}
\int\!{\rm u_1^2du_1}
\int\!{\rm u_2^2du_2}\nonumber\\
&&\exp(-i{ k_{1\mu}u_{1\mu}}+i{ k_{2\mu}u_{2\mu}})
\int\!{\cal D}z{\cal D}z^*\delta\left(z^*_\sigma z_\sigma
-4g^2\right)z^*_{\sigma_1}z_{\sigma_2}\nonumber\\
&&\times
\exp(-i\int\!\d^3x(j_\mu^{c_{01}}-j_\mu^{c_{02}})L^3_\mu)
\nonumber\\
&&\times\exp\left(i(k_{1\mu}-k_{2\mu})\epsilon_{\mu\nu\lambda}\partial_\nu
L^3_\lambda\right)
\exp\left(-\int\!{\rm d^3x} \,\partial_\mu z^*_\sigma
\partial_\mu z_\sigma\right)\nonumber \\
&&\int\!\D a\D b\, \exp\left( 4g^2\int\!\d^3x \,(2b_\mu L^3_\mu- b_\mu
b_\mu )\right)\nonumber \\
&&\times
\exp\left( {{i\theta }\over{4\pi^2}}
\int\!\d^3x\, (2a_\mu d_{\mu \lambda}L^3_\lambda
-a_\mu d_{\mu\lambda}
 a_\lambda )\right) ~.
\label{amubmu}
\end{eqnarray}
Where $c_{01}$ and $c_{02}$ are curves starting from some point $x_0$ and
ending at $x_1$ and $x_2$ respectively. We now make the shifts,
\begin{displaymath}
b_\mu\rightarrow b_\mu+\frac{i}{8g^2}(j^{c_{01}}-j^{c_{02}})~,
\nonumber
\end{displaymath}
and
\begin{displaymath}
a_\mu\rightarrow a_\mu-\frac{2\pi^2}{\theta}(\tilde k_{1\mu}-\tilde
k_{2\mu})~.
\nonumber
\end{displaymath}
The equation {\ref{amubmu}} can now be written as,
\begin{eqnarray*}
\langle\bar\chi_{\sigma_1u_1}\chi_{\sigma_2u_2}\rangle
&=&
\bar\zeta[c_{01}]\zeta[c_{02}]
\int\!\frac{\rm d^3k_1}{(2\pi)^3}
\int\!\frac{\rm d^3k_2}{(2\pi)^3}
\int\!{\rm u_1^2du_1}
\int\!{\rm u_2^2du_2}
e^{-i{ k_{1\mu}u_{1\mu}}+i{ k_{2\mu}u_{2\mu}}}\nonumber\\
&&\int\!{\cal D}z{\cal D}z^*\,\delta\left(z^*_\sigma z_\sigma-4g^2\right)
 z^*_{\sigma_1}z_{\sigma_2}\nonumber\\
&&\times
\int\!{\cal D}a{\cal D}b
e^{-\int\!{\rm d^3x\,}
\left(z_\sigma^*(i\partial_\mu-A_\mu)(i\partial_\mu-A_\mu)z_\sigma
\right)-S_{GF}}
\nonumber
\end{eqnarray*}
where we have dropped the terms that are zero for $x_1 \neq x_2$.
We now introduce the $\eta$ fields as before and do the
$z_\sigma$ integrals to get,
\begin{eqnarray}
\langle\bar\chi_{\sigma_1u_1}\chi_{\sigma_2u_2}\rangle
&=&
\bar\zeta[c_{01}]\zeta[c_{02}]
\int\!\frac{\rm d^3k_1}{(2\pi)^3}
\int\!\frac{\rm d^3k_2}{(2\pi)^3}
\int\!{\rm u_1^2du_1}
\int\!{\rm u_2^2du_2}\nonumber\\
&&\times
\exp(-i{ k_{1\mu}u_{1\mu}}+i{ k_{2\mu}u_{2\mu}})\nonumber\\
&&\times
\int\!{\cal D}a{\cal D}b\D\eta
\exp(-S_{GF}-S_\eta)\langle  x_2|\frac{1}{G}|x_1\rangle[\det G]^{-2}
\label{chichi}
\end{eqnarray}
apart from irrelevant constants.
The propogator $G^{-1}$ has a path integral representation as follows,
\begin{eqnarray*}
&&\langle  x_2|\frac{1}{G}|x_1\rangle
=
\langle  x_2|
\int_\frac{1}{\Lambda^2}^\infty   \!\d\beta\left(
\exp\left(-\beta G\right)
\right)|x_1\rangle\nonumber\\
&=&\int_0^\infty \!\d\beta
\int\!{\cal D}x\,e^{-\int_0^\beta{\rm d\tau}
( \frac{1}{4}\dot{x}^2-2i\eta\sqrt\lambda)}
e^{-i\int_{c_{12}} A_\mu{\rm dx_\mu}}
\nonumber
\end{eqnarray*}
where the path integral is over an open path $c_{12}$ from $x_1$ to
$x_2$.
For the determinant in (eq.{\ref{chichi}}), as was done earlier, we have,
\begin{eqnarray*}
[\det G]^{-2}&=&
\exp\left(-2\ln\det G\right)
=\sum_{n=0}^{\infty}
\frac{2^n}{n!}
\left(-{\rm tr}\ln G\right)^n
\end{eqnarray*}
where,
\begin{eqnarray*}
-\tr\ln G
=
\int_{1/{\Lambda^2}}^\infty \frac{\rm d\beta}{\beta}
\int\!{\cal D}x\,e^{-\int_0^\beta{\rm d\tau}
( \frac{1}{4}\dot{x}^2-2i\eta \sqrt\lambda)}
e^{-i\oint_c A_\mu{\rm dx_\mu}}
\end{eqnarray*}
where the path integral is over all closed loops.
Let us note that in both the above expressions the gauge field
dependence comes through Wilson loop integrals. The next step is to do
the gauge field integrals. For that we expand the ``$\det$'' term as a
series. Now if we collect all the Wilson loop terms a typical term would
be of the form,
\begin{displaymath}
\exp  \left(-i\int_{c_{12}} A_\mu\d x_\mu\right)
\exp \left(-i\left(\oint_{c_1}+\oint_{c_2}+\cdots+\oint_{c_n}\right)
A_\mu\d x_\mu\right)
\end{displaymath}
Let us note that except for the first integral (over $c_{12}$)
all other terms are
integrals over closed loops. Terms like these are to be substituted in
the expression for the correlation function and integrated over the
gauge fields. That is our next step. Collecting all the terms involving
$a_\mu$ and $b_\mu$ we get,
\begin{eqnarray}
&&\int\!{\cal D}a{\cal D}b
\exp\left(-\int\!\d^3x\left(4g^2(-2i\alpha b_\mu d_{\mu\lambda} a_\lambda
+\alpha^2(d_{\mu\lambda}a_\lambda)^2
+i\alpha a_\mu d_{\mu\lambda}  a_\lambda)
\right)\right)
\nonumber\\
&&\times
 \exp\left(-4g^2\int\!\d^3x \,\left(
 \frac{i}{4g^2}b_\mu(j_\mu^{c_{01}}-j^{c_{02}}_\mu)\right)\right)\nonumber\\
&&\times\exp\left(-{{i\theta }\over{4\pi^2}}
\int\!\d^3x \,\left(
-\frac{4\pi^2}{\theta}(\tilde k_{1\mu}-\tilde
k_{2\mu})
\epsilon_{\mu\nu\lambda}\partial_\nu a_\lambda
\right)\right)\nonumber\\
&&
\exp \left(-i\int_{c_{12}} A_\mu\d x_\mu\right)
\exp \left(-i\left(\oint_{c_1}+\oint{c_2}+\cdots+\oint_{c_n}\right)
A_\mu\d x_\mu\right)
\label{int}
\end{eqnarray}
Now let us introduce the following abbreviations. Let $c=\cup_i c_i$ and
$c_{012}= c_{01}\cup c_{12}\cup c_{20}$ where $c_{20} $ stands for $c_{02}$
traversed in the opposite direction. Not that $c_{012}$ stands for a
closed curve consisting of two fixed curves ($c_{01}$ and $c_{20}$)
coming from the ansatz and a ``fluctuating part'' coming from the path
integral for the matrix element of the propagator. With this the above
expression {\ref{int}} can be simplified to:
\begin{eqnarray*}
&&\int\!{\cal D}a{\cal D}b\,
\exp\left(-\int\!\d^3x\left(4g^2(-2i\alpha b_\mu d_{\mu\lambda} a_\lambda
+\alpha^2(d_{\mu\lambda}a_\lambda)^2
+i\alpha a_\mu d_{\mu\lambda}  a_\lambda)
\right)\right)
\nonumber\\
&&\times
 \exp\left(-i\int\!\d^3x \,\left(
b_\mu(j_\mu^{c_{01}}-j^{c_{02}}_\mu)\right)\right)
\exp\left(+i
\int\!\d^3x \,\left(
(\tilde k_{1\mu}-\tilde
k_{2\mu})
\epsilon_{\mu\nu\lambda}\partial_\nu a_\lambda
\right)\right)\nonumber\\
&&
\exp  \left(-i\int\!\d^3x j_\mu^{c_{12}} A_\mu\right)
\exp  \left(-i\int\!\d^3x j_\mu^{c} A_\mu\right)
\nonumber
\end{eqnarray*}
where
\begin{eqnarray*}
 j^c_\mu=\sum_i\int_{c_i}\d x^i_\mu\,\delta(x_\mu-x^i_\mu)\nonumber\\
j_\mu^{c_{012}}=j_\mu^{c_{01}}-j_\mu^{c_{02}}+j_\mu^{c_{12}}\nonumber\\
j_\mu^{c_{012}}=\int_{c_{012}}\d x^\prime_\mu\delta(x_\mu-x^\prime_\mu)
\nonumber
\end{eqnarray*}
Now doing the $b$ integrals (remembering that $A_\mu=b_\mu+i\alpha
d_{\mu\nu}a_\nu$)  we get,
\begin{eqnarray*}
&&\int\!{\cal D}a\,\delta\left(-8g^2\alpha d_{\mu\nu}a_\nu
-j^{c_{012}}_\mu-j^{c}_\mu\right)\nonumber\\
&&\times
\exp\left(-\int\!\d^3x\left(4g^2(
+\alpha^2(d_{\mu\lambda}a_\lambda)^2
+i\alpha a_\mu d_{\mu\lambda}  a_\lambda)
\right)\right)
\nonumber\\
&&\times\exp\left(i
\int\!\d^3x \,\left(
(\tilde k_{1\mu}-\tilde
k_{2\mu})
\epsilon_{\mu\nu\lambda}\partial_\nu a_\lambda
\right)\right)\nonumber\\
&&
\exp\left(\alpha \int\!\d^3x j_\mu^{c_{12}}  d_{\mu\nu}a_\nu \right)
\exp \left(\alpha \int\!\d^3x j_\mu^{c}
d_{\mu\nu}a_\nu \right)
\end{eqnarray*}
Now we can use the delta function connecting the `` a ''flux and the
particle current to do the $a_\mu$ integral. Solving the delta function
condition for $a_\mu$ we get,
\begin{displaymath}
d_{\mu\nu}a_\nu=\frac{2\pi^2}{\theta}(j_\mu^{c_{012}}+j^c_\mu)
\end{displaymath}
Doing the $a_\mu$ integral we get,
\begin{eqnarray*}
&&\exp\left(\frac{1}{16g^2}\int\!\d^3x
\left(j_\mu^{c}+j^{c_{12}}_\mu\right)^2\right)\nonumber\\
&&\exp\left(-\frac{i\pi^2}{\theta}\int\!\d^3x\,
(j_\mu^{c_{012}} +j^c_\mu)[d^{-1}]_{\mu\nu}(j_\nu^{c_{012}}+j^c_\nu) \right)
\nonumber\\
&&\times\exp\left(i
\int\!\d^3x \,\left(
(\tilde k_{1\mu}-\tilde
k_{2\mu})\frac{2\pi^2}{\theta}(j_\mu^{c_{012}}+j^c_\mu)
\right)\right)
\nonumber
\end{eqnarray*}
Of the terms above,
\begin{eqnarray*}
&&\exp\left(\frac{-i\pi^2}{\theta}\int\!\d^3x\,
(j_\mu^{c_{012}} +j^c_\mu)[d^{-1}]_{\mu\nu}(j_\nu^{c_{012}}+j^c_\nu) \right)
\nonumber\\
&&=(-1)^{2s}\exp(is\Omega[c_{012}])(-1)^{2s}\exp(is\Omega[c])
\exp\left(\frac{-i\pi^2}{\theta}2n[c_{012},c]\right)
\nonumber
\end{eqnarray*}
Here $n[c_{012},c]$ is the value of the Gauss integral for the curves
$c_{012}$ and $c$ and is equal to an integer. Thanks to this the last
term in the above expression can be dropped for the special values of
$\theta~=~\pi/2s$ that we have chosen. Also  by introducing an auxiliary field
$v$ we can write,
\begin{eqnarray*}
&&\exp\left(\frac{1}{16g^2}\int\!\d^3x
\left(j_\mu^{c}+j^{c_{12}}_\mu\right)^2\right)~~~~~~\nonumber\\
&=&\int\!\D v\exp\int\!\d^3x\left(-v_\mu v_\mu-\frac{1}{2g}
\left(j_\mu^{c}+j^{c_{12}}_\mu\right)v_\mu\right)
\nonumber
\end{eqnarray*}
As was argued earlier these ensure factorisation of Wilson loop
integrals.
So putting together all terms, we see that the effect of $a_\mu$ and
$b_\mu$ integrals is that,
\begin{displaymath}
\exp\left(-i\oint_c A_\mu\d x_\mu\right)
\rightarrow
(-1)^{2s}\exp\left(is\Omega[c]\right)
\end{displaymath}
So the inverse of $G$ avraged ovver the gauge fields becomes,
\begin{eqnarray*}
\langle  x_2|\frac{1}{G}|x_1\rangle
&\rightarrow&
\langle  x_2|
\int_\frac{1}{\Lambda^2}^\infty  \!\d\beta\left(
\exp\left(-\beta G\right)
\right)|x_1\rangle\nonumber\\
&=&\int_\frac{1}{\Lambda^2}^\infty  \!\d\beta\left(
\int\!{\cal D}x\,e^{-\int_0^\beta{\rm d\tau}
( \frac{1}{4}\dot{x}^2-2i\eta\sqrt\lambda)}
(-1)^{2s}e^{is\Omega[c_{012}]}\right.\nonumber\\
&&\left.\rule{0mm}{6mm}
e^{-\frac{1}{2g}\int\!\d^3x\,v_\mu j^{c_{12}}_\mu}
\right)~.
\nonumber
\end{eqnarray*}
and,
\begin{eqnarray*}
-\tr\ln G
&\rightarrow&
\int_\frac{1}{\Lambda^2}^\infty  \frac{\rm d\beta}{\beta}
\left(
\int\!{\cal D}x\,\exp\left(-\int_0^\beta{\rm d\tau}
( \frac{1}{4}\dot{x}^2-2i\eta \sqrt\lambda)\right)
(-1)^{2s}e^{is\Omega[c]}\right.\nonumber\\
&&-\left.\rule{0mm}{6mm}
\exp\left(-\frac{1}{2g}\int\!\d^3x\,v_\mu j^c_\mu\right)
\right)~.
\nonumber
\end{eqnarray*}

Now if we substitute all these in the expression  for the correlation
function (see Eq.(\ref{chichi})) the $k_1$, $k_2$, $u_1$, and $u_2$
integrals will give,
\begin{displaymath}
\Delta\left(u_{1\mu}-j_\mu^{c_{012}}(x_1)-j_\mu^c(x_1)\right)
\Delta\left(u_{2\mu}-j_\mu^{c_{012}}(x_2)-j_\mu^c(x_2)\right)
\nonumber
\end{displaymath}
If we assume the paths not to intersect
\begin{displaymath}
j_\mu^c(x_1)=
j_\mu^c(x_2)=0
\nonumber
\end{displaymath}
Also we can write,
\begin{displaymath}
\exp\left(is\Omega[c_{012}]\right)
=
\exp\left(is\Omega[c_{01}]\right)
\exp\left(is\Omega[c_{12}]\right)
\exp\left(-is\Omega[c_{02}]\right)
\nonumber
\end{displaymath}
Now let us consider:
\begin{eqnarray*}
&&\int_\frac{1}{\Lambda^2}^\infty  \!\d\beta
\int\!{\cal D}x
\Delta\left(u_{1\mu}-j_\mu^{c_{012}}(x_1)\right)
\Delta\left(u_{2\mu}-j_\mu^{c_{012}}(x_2)\right)\nonumber\\
&&\times \exp\left(-\int_0^\beta{\rm d\tau}
( \frac{1}{4}\dot{x}^2-2i\eta\sqrt\lambda)\right)
(-1)^{2s}e^{is\Omega[c_{012}]}
e^{-\frac{1}{2g}\int\!\d^3x\,v_\mu j^{c_{12}}_\mu}
{}~.
\nonumber
\end{eqnarray*}
If we now introduce velocity variables through delta functions etc. as
was done in appendix A, we will get this equal to
\begin{displaymath}
\langle x_2|\langle u_2|\frac{1}{\left( sgn(\theta)\frac{4}{\sqrt{\pi\epsilon}}
(i\partial_\mu-\frac{1}{2g}
v_\mu)\frac{T_\mu}{s}+\frac{\ln(2s+1)}{\epsilon}-2i\eta\sqrt\lambda
\right)}|u_1\rangle|x_1\rangle
\nonumber
\end{displaymath}
using the results proved earlier we have,

\underline{for integer $s$}
\begin{displaymath}
[\det G]^{-2}\rightarrow
\left(\det\left(
\frac{4}{\sqrt{\pi\epsilon}}sgn(\theta)
(i\partial_\mu-\frac{1}{2g}
v_\mu)\frac{T_\mu}{s}+\frac{\ln(2s+1)}{\epsilon} -2i\eta\sqrt\lambda
\right)\right)^{-2}
\nonumber
\end{displaymath}
and \underline{for half-odd integer $s$}
\begin{displaymath}
[\det G]^{-2}\rightarrow
\left(\det\left(\frac{4}{\sqrt{\pi\epsilon}}sgn(\theta)
(i\partial_\mu-\frac{1}{2g}
v_\mu)\frac{T_\mu}{s}+\frac{\ln(2s+1)}{\epsilon}-2i\eta\sqrt\lambda
\right)\right)^2
\nonumber
\end{displaymath}
So if we substitute everything in the expression for the $2$-point
correlation function we have:
\begin{eqnarray*}
&&\langle\bar\chi_{\sigma_1u_1}\chi_{\sigma_2u_2}\rangle
=
\bar\zeta[c_{01}]\zeta[c_{02}]
\exp\left(is\Omega[c_{01}]\right)
\exp\left(-is\Omega[c_{02}]\right)\nonumber\\
&&\int\D \eta\exp\left(-S_\eta\right)\exp\left(-\int\!\d^3x\,v_\mu
v_\mu\right)
\int\D\bar\psi\D\psi\,\bar\psi(x_1u_1)\psi(x_2 u_2)\nonumber\\
&&\exp\left(-\int\!\d^3x ~{\bar\psi}\left( sgn(\theta)
\frac{4}{\sqrt{\pi\epsilon}}
(i\partial_\mu-\frac{1}{2g}
v_\mu)\frac{T_\mu}{s}+\frac{\ln(2s+1)}{\epsilon}-2i\eta\sqrt\lambda
\right)\psi\right)
\nonumber
\end{eqnarray*}
Choosing $\zeta[c_{01}]~\equiv (sgn(\theta){\sqrt{\pi \epsilon} \over 4})^{1
\over 2}
exp(is\Omega [c_{01}])$, redefining $\psi$ to absorb $sgn(\theta){4
\over\sqrt{\pi
\epsilon}}$ in the kinetic energy term and doing the $v_\mu$ and $\eta$
integrals, we obtain our result,
\begin{equation}
\langle \bar\chi_{\sigma_1 u_1}(x_1) \chi_{\sigma_2 u_2}(x_2)
{}~=~
\int\D\bar\psi\psi~\bar\psi_{\sigma_1 u_1}(x_1) \psi_{\sigma_2 u_2}(x_2)
e^{-S_\psi[\bar\psi,\psi]}
\label{2corr}
\end{equation}
Where $S_\psi$ is the spin-s theory action defined earlier in equation
\ref{hs}. Thus we have established that the two point correlation
functions of the $\chi$ operators in the NLSM as defined in equation
\ref{chibarchi} are equal to the two point functions of the $\psi$ operators in
the spin-s theory.
\vskip.75cm
\noindent{\bf 6. Current Correlators}
\vskip.75cm
Having established the equivalence of two point functions now
we shall  derive certain relationships between the
current-current correlation functions in the topological current in the
NlSM and the Noether current in the spin-$s$ theories. To do this we will
consider the partition function of the NLSM with an external gauge field
coupled to the topological current, i.e. the generating functional for
the topological current correlators. In the spin-s theory representation,
it turns out to be related to the generating functional for the Noether
current correlators. In this way we are able to relate the two
correlators.

We have,
\begin{displaymath}
j^{top}_\mu(x)
\equiv\frac{1}{2\pi}
\epsilon_{\mu\nu\lambda}
\partial_\nu L^3_\lambda~.
\end{displaymath}
Now we couple an auxiliary field $c_\mu$ to the topological current and
write down the generating functional for the topological current
correlators,
\begin{equation}
Z[c_\mu]
{}~=~
\int\!\D z\D z^*
e^{-S_{CP_1}+\frac{1}{2\pi}\int\!\d^3x\,c_\mu
\epsilon_{\mu\nu\lambda}
\partial_\nu L^3_\lambda
}\label{nlsmc}.
\end{equation}
Now introduce the auxilliary $a_\mu$ and $b_\mu$ fields as before and
make the  change of variable,
\begin{displaymath}
a_\mu\rightarrow a_\mu-\frac{2\pi^2}{i\theta}c_\mu~.
\end{displaymath}
we get,
\begin{displaymath}
Z[c_\mu]
{}~=~
\int\!\D z\D z^*\D a\D b
e^{-\tilde{S}_{CP_1}}
 e^{-\frac{i\theta}{4\pi^2}\int\!\d^3x
\,\left(
-\frac{2\pi}{i\theta}
\epsilon_{\mu\nu\lambda}
a_\mu\partial_\nu c_\lambda
-(\frac{\pi}{\theta})^2
\epsilon_{\mu\nu\lambda}
c_\mu\partial_\nu c_\lambda
\right)}
\end{displaymath}
Now we  can repeat the procedure in section 5. and get,
\begin{equation}
\label{psic}
Z[c_\mu]~=~
\int\D\bar\psi\D\psi~ e^{-S_\psi[\bar\psi,\psi]+
\int{\rm d^3x}\left( 2sj^N_\mu c_\mu
+i({s \over \pi})^2\epsilon_{\mu\nu\lambda}
c_\mu\partial_\nu c_\lambda
\right)}
\end{equation}

We can now relate the current correlators by differentiating equations
\ref{nlsmc} and \ref{psic}  with respect to $c_\mu$ and then evaluating at
$c_\mu~=~0$. By differentiating once we get,
\begin{equation}
\label{1curr}
\langle j_\mu^{top}(x) \rangle ~=~2s\langle j_\mu^N(x)\rangle
\end{equation}
Thus we see that the soliton current is proportional to the the Noether
current of spin-s theory. The relation also tells us that the spin-s
particle corresponds to a soliton number 2s object. This is consistent
with the spin of the particle being ${1 \over 2}(2s)^2{1 \over 2s} ~=~ s$
as expected from the NLSM for a soliton number 2s object (see equation
\ref{nsol}). Similarly the statistics is also consistent. This relation
also tells us that if the $\psi$ particles were charged then the charge of the
solitons is ${1 \over 2s}$ times the charge of a particle.

Next we can differentiate twice and obtain the following relation between
the current-current correlators,
\begin{equation}
\label{2curr}
\langle j_\mu^{top}(x)j_\nu^{top}(y)\rangle~=~
(2s)^2\langle j_\mu^N(x)j_\nu^N(y)\rangle~+~
i{s \over \pi}\epsilon_{\mu\nu\lambda}\partial_{\lambda}\delta^3(x-y)
\end{equation}
This then shows that the two currents are not exactly the same but differ
in correlation functions involving products of currents at the same
space-time points. The significance of the above relation is discussed
later.
% This file contains
% Lorentz and discrete transformations  of the ansatz for  spin-$s$
% fields in terms of $CP_1$ fields

\def \d {{\rm d}}
\def \c {{\cal C}}
\def \p {{\prime }}
\vskip.75cm
\noindent{\bf 7. Lorentz and discrete transformations:}
\vskip.75cm
In this section we will show that our ansatz for spin-$s$ fields has the
correct
transformation properties under Lorentz and discrete tranformations. We
will do this by deriving the transformation properties of the correlation
functions of the $\chi$ fields and showing that it is the same as the
transformation properties of the correlation functions of the spin-s
fields. The transformation properties of the spin-s fields in the
coherent state basis that we are using is not standard. We have therefore
discussed them in detail in Appendix B.

\noindent{\underline{1. Proper Lorentz rotations}} \\
In a Lorentz rotated frame,
we would have
$ x^\prime=\Lambda x~~~{\rm and}~~~
u^\prime=\Lambda u$.
This implies
\begin{displaymath}
c^\prime=\Lambda c~~~{\rm and}~~~
\c^\prime=\Lambda \c\nonumber\\
\end{displaymath}
where $ \Lambda c $ is the curve obtained by making Lorentz
transformation to every point on $c$. Under Lorentz transformations we
also have,
\begin{eqnarray*}
z^\prime_\sigma(x^\prime)&=&z_\sigma(x)\nonumber\\
L^{3\prime}_\mu(x^\prime)&=&(\Lambda^{-1}L^3(x))_\mu\nonumber\\
(dL^3)^\prime_\mu(x^\prime)&=&(\Lambda dL^3(x))_\mu
\end{eqnarray*}
 The two point function in the Lorentz rotated frame is
\begin{displaymath}
\langle\chi^\prime_{u^{\prime}\sigma^{\prime}}[x^{\prime},c^{\prime}]
{\bar \chi}^{\prime}_{{\bar u}^{\prime}{\bar \sigma}^{\prime}}
[{\bar x}^{\prime}, {\bar c}^{\prime}]\rangle.
\end{displaymath}
Now we have
\begin{displaymath}
\Omega[\c^{\prime},u^{*\prime}] = \Omega[\Lambda\c,\Lambda u^*] =
\Omega[\c,u^*].
\end{displaymath}
Also we have
\begin{eqnarray*}
\int_{c^\p} L^{3\p}_\mu dx^\p_\mu&=& \int_c L^3_\mu dx_\mu
\nonumber\\
\Delta \left(u^\p - (dL^{3\p}(x^\p))\right)& =& \Delta \left(\Lambda
u - \Lambda dL^3(x)\right) = \Delta \left(u - (dL^3(x))\right)
\end{eqnarray*}
and
\begin{displaymath}
z^\p_\sigma(x^\p) = z_\sigma(x) ~~~~~ z^*_\sigma({ x^\p}) = z_\sigma
({\bar x}).
\end{displaymath}
The action is invariant and
\begin{displaymath}
\langle \chi^\p_{u^\p \sigma^\p}[x^\p c^\p]{\bar \chi}^\p_{{\bar
u}^\p{\bar \sigma}^\p}[{\bar x}^\p {\bar c}^\p]\rangle = \langle\chi_{
u\sigma}[x,c]{\bar \chi}_{{\bar u}{\bar \sigma}}[{\bar x},{\bar c}].
\rangle
\end{displaymath}
This then implies,
\begin{displaymath}
\langle \psi_{u^\p\sigma}(x^\p){\bar \psi}_{{\bar u}^\p{\bar \sigma}}
({\bar x}^\p)\rangle = \langle \psi_{u\sigma}(x){\bar \psi}_{{\bar u}{
\bar \sigma}}({\bar x})\rangle.
\end{displaymath}
This is the correct transformation of the spin-s fields under proper
Lorentz transformations as has beeen shown in equation (B.7).

\noindent{\underline{2. Parity}}

In a parity transformed frame, we have
\begin{eqnarray*}
x^\p = -x~~~u^\p = -u~~~c^\p = -c~~~\c^\p = -\c\\
z^\p_\sigma(x^\p) = z_\sigma(x) ~~~~z^{*\p}_\sigma(x^\p) =
z^*_\sigma(x)\\
L^{3\p}_\mu(x^\p) = -L^3_\mu(x) ~~~~dL^{3\p}_\mu(x^\p) = dL^3_\mu(x)
\end{eqnarray*}
But now the action is not invariant since the Hopf term changes sign.
Thus in the parity transformed frame we have,
$$ \theta '~=~-\theta~=~-{\pi \over 2s}~=~{\pi \over 2s'}$$
This then implies that,
\begin{eqnarray*}
s^\p\Omega[\c^\p,u^{*\p}]& =& (-s)\Omega[-\c,-u^*] = s\Omega[\c,u^*]\nonumber\\
\Delta\left(u^\p - dL^{3\p}(x^\p)\right)& =& \Delta \left((-u)-dL^3(x)
\right)\nonumber\\
\int_{c^\p} L^{3\p}_\mu dx^\mu& =& \int_c L^3_\mu dx^\mu\nonumber\\
\sqrt {sgn(\theta ')}~&=&~i\sqrt{sgn(\theta)}
\end{eqnarray*}
We then have,
\begin{displaymath}
\langle\chi^\p_{u^\p\sigma}[x^\p,c^\p]{\bar \chi}^\p_{{\bar u}^\p\sigma}
[{\bar x}^\p,{\bar c}^\p]\rangle_{s^\p} =
-\langle\chi_{-u^\p\sigma}[x,c]{\bar \chi}_{-{\bar u}\sigma}[{\bar
x},c]\rangle_s
\end{displaymath}
This then implies
\begin{displaymath}
\langle \psi^\p_{u^\p\sigma}[x^\p]{\bar \psi}^\p_{{\bar
u}^\p\sigma}[{\bar x}^\p]\rangle_{s^\p_F} =
-\langle\psi_{-u^\p\sigma}[x]{\bar \psi}_{-{\bar u}\sigma}[{\bar
x}]\rangle_{s_F}
\end{displaymath}
where the subscripts refer to the action. $ S^\p_F $ differs from $ S_F
$ in the sign of the mass term. This as shown in equation (B.8) is the
correct parity transformation of the propogator.

\noindent{\underline {3. Charge Conjugation}}

Under charge conjugation we have,
\begin{displaymath}
x^\p = x ~~~{\rm and}~~~c^\p = c^{-1}
\end{displaymath}
where $ c^{-1} $ denotes the same path traversed in the opposite sense.
This is because $ c $ is physically the world line of a particle and
will thus reverse under charge conjugation. That is,
\begin{displaymath}
u^\p = -u,~~\c^\p = -\c^{-1}
,~~~z^\p_\sigma(x^\p) = z^*_\sigma(x) ~~~z^{*\p}_\sigma
(x^\p) = z_\sigma(x),
\end{displaymath}
which imply that,
\begin{displaymath}
L_\mu^{3\p}(x^\p)
 = -L^3_\mu(x) ~~~{\rm and}~~~ dL^{3\p}_\mu(x^\p) = -dL^3_\mu(x).
\end{displaymath}
Let us note that the action is invariant under this transformation.

We now have,
\begin{eqnarray*}
\Omega[\c^\p,u^{*\p}] = \Omega[-\c^{-1},-u^*] = \Omega[\c,u^*]\\
\Delta\left(u^\p- dL^{3\p}_\mu(x^\p)\right) = \Delta \left(u-dL^3_\mu
(x)\right)
\end{eqnarray*}
With these we have,
\begin{displaymath}
\langle\chi^\p_{u^\p\sigma}[x^\p,c^\p]{\bar \chi}^\p_{{\bar u}^\p\sigma}
[{\bar x},{\bar c}]\rangle =
\langle\chi_{{\bar u}\sigma}[{\bar x},{\bar c}^{-1}]{\bar \chi}^\p_{{u}\sigma}
[{x},c^{-1}]\rangle
\end{displaymath}
Note that $ {\bar c}^{-1} $ starts at $ \infty $ and ends at $ {\bar x}
$ and $ c^{-1} $ starts at $ x $ and ends at $ \infty $. So we have,
\begin{displaymath}
\langle\psi^\p_{u^\p\sigma}(x^\p){\bar \psi}^\p_{{\bar
u}^\p\sigma}(x)\rangle = \langle\psi_{{\bar u}\sigma}({\bar x}){\bar
\psi}_{u\sigma}(x)\rangle
\end{displaymath}
This, as shown in equation (B.9) is the correct transformation under
charge conjugation.
\vskip.75cm
\noindent {\bf 8.Discussion and Conclusions:}
\vskip.75cm
To summarize, using the $CP_1$ formalism, we have shown that the
z-theory defined in eq.(\ref{z}) is formally exactly equivalent to the NLSM,
in the limit $\lambda \rightarrow \infty$. Then we showed that the
partition function of the z-theory is exactly equivalent to the
partition funtion of the spin-s theory defined in eq.(\ref{hs}). Next we
constructed nonlocal operators in the NLSM and showed that the
correlation functions of these operators in the z-theory are exactly
equal to the correlation functions of the $\psi$ operators in the spin-s
theory. This established the exact equivalence between the z-theory and
the spin-s theory.
The above result then strongly suggests that there should be a
nontrivial $\lambda = \infty$ fixed point of the spin-s theory (and
hence of the z-theory) near which the continuum theory should be exactly
equivalent to the NLSM.

We now give some physical arguments for the plausibility of the
existence of such a fixed point. The spin-s theory at large $\lambda$ is
a system of strongly interacting particles in two dimensions. For
non-zero $M_s$, the mass term, ${\bar \psi} \psi$, breaks parity.
Fractional Quantum Hall Effect (FQHE) systems are analagous
non-relativistic systems. There again we have a system of strongly
interacting particles in two dimensions where the interaction with the
external magnetic field breaks parity. There are very good approximate
solutions of the FQHE system \cite{FQHE} fom which we know that the
system is characterised by fractionally charged anyonic excitations.
These excitations are soliton like, in the sense that they are extended,
finite energy objects. Now this exactly the situation suggested by our
result. Namely that the system of strongly interacting spin-s particles
is equivalent to the NLSM in the continuum limit which has anyonic
soliton excitations.

The analogy to FQHE systems can be made more precise following a very
general analysis of such systems, dubbed as Quantum Hall fluids, by
Frolich and Zee \cite{FZ}. They classify these systems into universality
classes characterised by three universal properties. (i) The Hall
conductivity, $\sigma_H$. (ii) The charge of the excitations, q. (iii)
The statistics of the excitations, $\theta \over \pi$. They have shown
that, in the simplest case \cite{FZ}, the allowed values of these
quantities are,
\begin{equation}
\sigma_H~=~{1 \over p+1}
\label{qha}
\end{equation}
\begin{equation}
q~=~{1 \over p+1} \Phi
\label{qhb}
\end{equation}
\begin{equation}
{\theta \over \pi}~=~{\Phi^2 \over p+1}
\label{qhc}
\end{equation}
Where p+1 is an odd integer if the underlying particles are fermions an
even integer if they are bosons. $\Phi$ is the vorticity of the
excitations. We will now show that the NLSM can be looked upon as a
Quantum Hall fluid of the above type with p+1=s, when the soliton number
is identified with the vorticity $\Phi$.

The fact that the statistics of a soliton number $n_s$ object is
$n_s^2{\pi \over 2s}$ has already been mentioned in equation (\ref{nsol}).
 Thus our identifications are consistent with equation (\ref{qhc}). Next,
from the relation between the spin-s Noether current and the soliton
current (\ref{1curr}, as mentioned before, the charge of the soliton is ${1
\over
2s}$ if the charge of the spin-s particle is one. Thus the charge of the
excitations is consistent with equation (\ref{qhb}). Finally we come to
the Hall conductivity. $\sigma_H$ is related to the current-current
correlator as follows,
\begin{equation}
\sigma_H~=~
\lim_{\omega \rightarrow 0}{1 \over \omega}\epsilon_{ij}
<J_i(\omega,0)J_j(-\omega,0)>
\label{sigh}
\end{equation}
Note that here as in reference (\cite{FZ}), by $\sigma_H$ we mean the
off-diagonal components of the conductivity tensor at {\it zero}
external field. From equation (\ref{2curr}), it follows that,
\begin{displaymath}
\sigma_H~=~{1 \over 2s}~+~\lim_{\omega \rightarrow 0} {1 \over
\omega}\epsilon_{ij}<j^{top}_i(\omega,0)j^{top}_j(-\omega,0)>
\nonumber
\end{displaymath}
We will now argue that the second term in the RHS of the above equation
is zero. This we will do by arguing that if we couple an external field
to the toplogical current in the NLSM, then the induced Chern-Simons
term will be zero. Consider the NLSM action in equation (\ref{nlsmc}).
Now introduce the auxilliary $a_\mu$ fields as before to obtain the action,
\begin{equation}
\label{nlsmca}
S[\hat{n},a_\mu,c_\mu]~=~\int d^3x~{1 \over g^2}\partial_\mu \hat n
\partial_\mu \hat n~+~j_\mu^{top}(2i\theta a_\mu+c_\mu)~+~i\theta
\epsilon_{\mu \nu \lambda} a_\mu \partial_\nu a_\lambda
\end{equation}
Now integrate over the $\hat n$ fields to obtain the following form of
the action,
\begin{equation}
\label{Fa}
S[a_\mu, c_\mu]~=~F[2i\theta a_\mu+c_\mu]~+~i\theta\epsilon_{\mu \nu \lambda}
a_\mu \partial_\nu a_\lambda
\end{equation}
Now note that the part of the action in equation (\ref{nlsmca}) that
involves only the $\hat n$ fields is parity invariant (parity is broken
only by the Chern-Simons term). Therefore F will be a parity invariant
functional and cannot contain any terms like $\epsilon_{\mu \nu \lambda}
(2i\theta a_\mu + A_\mu) \partial_\nu (2i\theta a_\lambda + A_\lambda)$.
The action in equation (\ref{Fa}) is then independent of $c_\mu$ in the
long wavelength limit. Hence there is no induced Chern-Simons term when
the $a_\mu$ fields are integrated out. It then follows that we have,
$\sigma_H~=~{1 \over 2s}$. Thus our identifications are consistent with
equation (\ref{qha}). This then establishes the fact that the NLSM is a
Quantum Hall fluid of the type specified by equations (\ref{qha}),
(\ref{qhb}) and (\ref{qhc}).

The equivalence of the spin-s theory and the NLSM is very convincing
when viewed in the light of the above arguements. Namely, that the
strongly interacting, parity non-invariant spin-s system should have a
Quantum Hall Fluid phase, characterised by some universal properties.
The corresponding NLSM, along with the relations between the current
correlators that we have derived, is consistent with the general
analysis of these universal properties done in ref.\cite{FZ}. Also,
we have shown the equivalence of the two theories in the formal $\lambda
\rightarrow \infty$ limit. Therefore the spin-s theory should have a
$\lambda~=~\infty$ fixed point which governs this phase of the system.
However the $\theta=0~and ~\pi$ limits (where the NLSM is parity
invariant) are likely to be singular just
like $B~=~0$ limit in Quantum Hall systems\cite{CAN}.These two values
of $\theta$ are  thus beyond the purview of the above discussion. Note
that we had previously argued for different reasons that the $\theta=0$
limit is likely to be singular (see the discussion after equation
(\ref{jflux})). However the $\theta~=~\pi$ point is different. Here our
results indicate that the theory should be equivalent to a system of
stongly interacting Dirac fermions though it is not a Quantum Hall fluid.
Since the NLSM is parity invariant at this point, the mass term in the
spin-$1 \over 2$ theory should become an irrelevent operator at the
$\lambda~=~\infty$ fixed point so that the continuum theory is parity
invariant. This also shows that the usual large N expansion \cite{ROS}
will not see this fixed point. This is understandable since it
corresponds to generalizing $CP_1$ to $CP_{N-1}$. This spoils the
topological features of the N = 2 model which are crucial for the parity
invariance. Finding a good expansion (that preserves the topology) to
locate and analyse this fixed point is clearly an important problem.
\vskip.75cm
\noindent {\bf Appendix A:}
\vskip.75cm
\noindent {\bf Sum over path representation of spin-$s$ particle in external
fields :}

It is now well-known \cite{AL} that addition of Polyakov spin factor to path
integrals,
for free spinless particles gives path integral representation for spinning
particles. We now show that this proof can be extended to particle in the
presence of background fields.

We start with the sum over path representation for spinless particle with
spin-factor coupled to
 vector and scalar fields,
$$
K_{A,V} (x,x|\beta) \equiv \int_{x(o)=x(\beta)} Dx \ e^{-\int^\beta_0
\frac{(\partial_\tau x)^2}{4}~+~V(x(\tau))+
{}~is\Omega [s] d\tau~-~i\oint A_\mu dx^\mu} \eqno(A.1)
$$
Here $K(x,x|\beta)_{A,V}$ is the amplitude for the particle to make a closed
curve in time $\beta$.

We show that this is equal to,
$$
Tr \ e^{- L (D^{(s)} + \frac{\sqrt{\pi}\epsilon}{4}V(x) + M_s)} \eqno(A.2)
$$
where
$$
L \ = \ \frac{\beta}{\kappa} \quad M_s \ = \ \Lambda^2 \kappa~ ln(2s+1)
$$
and
$$
D^{(s)} \ = \ sgn(\theta) \left( i \partial_\mu + A_\mu \right) \frac{T^\mu}{s}
$$
with $T^\mu$ being the generators of $SU(2)$ in spin-$s$ representation.

Starting with (A.1), we first change the variable in the integrand from
$\frac{dx^\mu}{d\tau}$ to $u^\mu(\tau)$, using the identity,
$$
F[x,\dot{x}] \ = \ \int Du \delta^{(3)} [\dot{x}-u] F[x,u] \eqno(A.3)
$$

\noindent $K(x,x|\beta)$
$$
= \int \prod_\tau \frac{d^3 x(\tau)}{(4\pi \epsilon)^{3/2}} \int d^3 u(\tau)
\delta^3 (\dot{x}_\mu (\tau) - u_\mu (\tau))e^{ \int \left[ \frac{u^2}{4} - i
u.A +V[x]\right] d\tau -is \int \Omega [\hat{u}]} \eqno(A.4)
$$
where $\epsilon$ is the interval between two adjacent $\tau$-slices.
Replace $\prod_\tau \delta^3 (\dot{x} -u)$ by
$$
\prod_\tau \int \frac{d^3k(\tau)}{(2\pi)^3} \epsilon^3e^{i\left[k \cdot
(x(\tau +1) - x(\tau) - \epsilon k \cdot u(\tau) \right]} \eqno(A.5)
$$
The path-integral over the velocity variables are performed in radial
and polar variables. The need for doing it this way will be seen later. The
radial integral to be performed is,
$$
\int^\infty_0 u^2 du e^{- i \epsilon u \hat{u} \cdot \vec{K} - \epsilon
\frac{u^2}{4}}
$$
where $\vec{K} \ = \ \vec{k} - \vec{A}$
Scaling $u_\mu \rightarrow \sqrt{\epsilon} u_\mu$ and keeping only
$O(\sqrt{\epsilon})$ term, this becomes,
$$
\frac{2 \sqrt{\pi}}{\epsilon^{3/2}} \  e^{-i4\sqrt{\frac{t}{\pi}}
\hat{u} . \vec{K}} \eqno(A.6)
$$
With this the measure for the remaining path integral to be
performed over $\vec{x}, \vec{k}$ and $\vec{u}$, are
$$
\prod_\tau \frac{d^3 x(\tau)}{4\pi} \ \frac{d^3 k(\tau)}{(2\pi)^3} \ d\hat{u}
$$
The path-integral to be performed is,
$$
\prod^N_{\tau = 1} \int d^3 x(\tau) \ e^{u(x)} \int \frac{d^3k(\tau)}{(2\pi)^3}
\int \frac{d\hat{u}}{4\pi}e^{\sum^N_{i=1} \left\{iw^2
\epsilon \left[ \hat{u} \cdot \vec{K}(\tau)\right] + is
\Omega[\hat{u}]\right\}}
\eqno(A.7)
$$
where $w^2 \ = \ \frac{4}{\sqrt{\pi}\epsilon}$
The need for performing the path integral over $u^\mu$ in radial and angular
variables is now clear : with the Polyakov spin-factor $\Omega [\hat{u}]$,
which is the solid angle subtended by the closed curve on the sphere $S^2$,
traced by the unit velocity vector, having the geometrical meaning of providing
the symplettic structure of $SU(2)$ group the path integral over unit vector
$\hat{u}$ (with measure $\frac{d \hat{u}}{4\pi} (2s+1)$, is the phase-space
path integral for $SU(2)$ group.

In general,
$$
\int_{\hat u(0)=\hat u(s)}{d \hat{u}}\frac{(2s+1)}{4\pi}e^{is \int^s_0 \left[
\Omega [\hat{u} ] + H[ \hat{u} ] \right] d \tau}
$$
$$
=~~~~~ Tr<\hat{u} | e^{iH[\frac{T}{\beta}]\beta}|\hat{u}>
$$
where $\vec{T}$ are the generators in the spin-$s$ representation of $SU(2)$
group and
$|\hat{u}>$ is the $SU(2)$ coherent state.

This relation can be proved in a straight forward way using the following
properties of $SU(2)$ coherent states:

\vspace{.5cm}
\begin{eqnarray*}
\int\frac{2s+1}{4\pi}\int\d u| u u^*\rangle\langle
 u u^*|&=&I\!\!I~,\nonumber\\
\langle uu^*|\gamma^\mu|uu^*\rangle&=&isu^\mu~,\nonumber\\
\langle u_1u^*|u_2u^*\rangle&=&e^{is\Omega[u_1,u_2,u^*]}\left(
\frac{1+u_1\cdot u_2}{2}\right)^s~.
\end{eqnarray*}
where $u^*$ refers to any fiducial point on $S^2$.
%
%\noindent (1) \qquad $\int \frac{d\hat{u}}{4\pi} |\hat{u}> <\hat{u}|\ = \
%\frac{1}{2s+1}$
%
%\noindent (2) \qquad\qquad ~$<\hat{u}|\hat{u}> \ = \ e^{is\Omega
%[\hat{u}]}$ \hfill {(A.8)}
%
%\noindent (3) \qquad~~ $<\hat{u}|T^a|\hat{u}> \ = \ s \hat{u}^a$

\noindent Using as the overlap between states $|x(\tau)\hat{u}(\tau)>$ at
$\tau$ and $\tau +1$ namely \\
$\int\frac{d^3 k(\tau)}{(2\pi)^3}
\frac{d\hat{u}(\tau)}{4\pi} (2s+1)e^{\left[ik\cdot (x(\tau+1) -
x(\tau)) + is \Omega [\hat{u}] \right]}$ it follows easily, (A.7) is
$$
Tr<\hat{u}|e^{-sw^2\left[\frac{ln(2s+1)}{\epsilon w^2}-
\left(D + \frac{V(x)}{w^2} \right) \right]}|\hat{u}>.\eqno(A.8)
$$
Note that when $\theta ~=~-{\pi \over 2s}$, then in all equations the
solid angle term comes with a minus sign. To use the same conventions for
the definitions of the coherent states, we make the change of variables
$\hat{u}~\rightarrow ~ -\hat{u}$ in equation (A.7) since $\Omega
[-\hat{u}]~=~-\Omega[\hat{u}]$. This results in $K~\rightarrow~-K$ which
leads to $D^{(s)}~\rightarrow~-D^{(s)}$. Therefore, in general, the
operator that appears in equation (A.8) is,
$D^{(s)}~=~sgn(\theta )(i\partial_\mu - A_\mu ){T^\mu \over s}$.
 Thus with the definition $L \ = \ \beta w^2, \frac{1}{\epsilon w^2}ln(2s+1)~=
{}~M_s$ the claimed result (A.2) follows.

\vskip.75cm
\noindent{\bf Appendix B:}
\vskip.75cm
% This file contains
\noindent{\bf Lorentz and Discrete Transformation Properties
 of spin-$s$ Fields.}

\def \d {\rm d}
\def \c {{\cal C}}
\def \p {\prime }

%Spin-$s$ coherent states are defined by,
%\begin{eqnarray*}
%\int\frac{2s+1}{4\pi}\int\d u| u u^*\rangle\langle
% u u^*|=I\!\!I~,\nonumber\\
%\langle uu^*|\gamma^\mu|uu^*\rangle=isu^\mu~,\nonumber\\
%\langle u_1u^*|u_2u^*\rangle=e^{is\Omega[u_1,u_2,u^*]}\left(
%\frac{1+u_1\cdot u_2}{2}\right)^s~.
%\end{eqnarray*}
%Also
In this appendix we derive the transformation properties of the $\psi$
spin-s fields under Lorentz and the discrete transformations in the
coherent state representation. i.e.,
\begin{eqnarray*}
\psi_u(x)\equiv\langle uu^*|m\rangle \psi_m(x)\nonumber\\
\bar\psi_u(x)\equiv\bar\psi_m(x)\langle m|uu^*\rangle
\end{eqnarray*}

\noindent {\underline{1. Proper Lorentz rotations.}}

Under Lorentz transformations given by,
$$x^\p=\Lambda x
$$
we have
\begin{eqnarray*}
\psi^\p(x^\p)=U^\dagger(\Lambda)\psi(x)~,~~~
\bar\psi^\p(x^\p)=\bar\psi(x)U(\Lambda)~,\nonumber\\
U^\dagger(\Lambda)\gamma^\mu U(\Lambda)
=\gamma^\nu\Lambda_{\nu\mu}~~~{\rm and}
{}~~~U(\Lambda)\gamma^\mu U^\dagger(\Lambda)
=\Lambda_{\mu\nu}\gamma^\nu~.
\end{eqnarray*}
Therefore in the coherent state basis we have,
\begin{eqnarray*}
\psi_u(x)&=&
\langle uu^*|m\rangle \psi_m(x)\nonumber\\
&=&\langle uu^*|m\rangle U_{mm^\p}(\Lambda)\psi^\p_{m^\p}(x^\p)
\nonumber\\
&=&\langle uu^*|U(\Lambda)|m\rangle \psi^\p_{m^\p}(x^\p)~.
\end{eqnarray*}
Now we will prove that,
$$
U^\dagger(\Lambda)|uu^*\rangle
=
|\Lambda u,\Lambda u^*\rangle~.
$$
Therefore we have $\psi^\p_{u^\p}(x^\p)=\psi_u(x)$ where $u^\p=\Lambda
u$.

\noindent Proof of $U^\dagger(\Lambda)|uu^*\rangle=|\Lambda u,\Lambda u^*
\rangle$:

We have,
\begin{eqnarray*}
\frac{2s+1}{4\pi}\int\d u
U^\dagger(\Lambda)| u u^*\rangle\langle
 u u^*|U(\Lambda)&=&U^\dagger I\!\!IU(\Lambda)
=I\!\!I\nonumber~,\\
\langle uu^*|U(\Lambda)\gamma^\mu U^\dagger(\Lambda)
|uu^*\rangle&=&\Lambda_{\mu\nu}\langle uu^*|\gamma^\mu|uu^*\rangle
\nonumber\\
&=&\Lambda_{\mu\nu}\,is\,u^\nu=is(\Lambda u)^\mu~,\nonumber\\
\langle u_1u^*|U(\Lambda)U^\dagger(\Lambda)|u_2u^*\rangle
&=&e^{is\Omega[u_1u_2u^*]}\left(\frac{1+u_1\cdot u_2}{2}\right)^s~.
\end{eqnarray*}
But we have,
 $u_1\cdot u_2=(\Lambda u_1)\cdot (\Lambda u_2)$ and $\Omega[u_1u_2u^*]
=\Omega[\Lambda u_1,\Lambda u_2,\Lambda u^*]$ so that we can write,
\begin{eqnarray*}
\left(
\langle u_1u^*|U(\Lambda)\right)
\left(U^\dagger(\Lambda)|u_2u^*\rangle\right)
=e^{is\Omega[\Lambda u_1,\Lambda u_2,\Lambda u^*]}
\left(\frac{1+\Lambda u_1\cdot \Lambda u_2}{2}\right)^s
\end{eqnarray*}
That proves that $U^\dagger(\Lambda)|u,u^*\rangle =
|\Lambda u,\Lambda u^*\rangle$.

\noindent {\underline{2. Parity}}

Under $x^\p=-x$ we have $\psi^\p(x^\p)=i\psi(x)$ and
 $\bar\psi^\p(x^\p)=i\bar\psi(x)$. This implies that
$\psi^\p_u(x^\p)=i\psi_u(x)$ and
$\bar\psi^\p_u(x^\p)=i\bar\psi_u(x)$.

\noindent {\underline{3. Charge conjugation}}

Here we need to distinguish between the cases of $s$ being integer and
half-odd integer. First we will consider the case of integer $s$ and
then that of half-odd integer $s$.

\noindent{\bf Case i) s=integer:}

We have under charge conjugation $x^\p=x$,
$\psi^\p_m(x^\p)=\bar\psi_m(x)$ and $\bar\psi^\p_m(x^\p)=\psi_m(x)$
. With these we get,
$$
\psi_u(x)=\langle\,uu^*|m\rangle\bar\psi^\p_m(x^\p)
\equiv \bar\psi^\p_m (x^\p)\langle m|U\rangle
$$
and
$$
\bar\psi_u(x)=\bar\psi_m\langle m|u\,u^*\rangle
=\langle U|m\rangle \psi^\p_m(x^\p)
$$
where $\langle m|U\rangle\equiv \langle u\,u^*|m\rangle$.
We will show that, $|U\rangle=|-u,-u^*\rangle$. We then have
\begin{eqnarray*}
\psi^\p_{u^\p}(x^\p)=\bar\psi_u(x)~~~x^\p=x\nonumber\\
\bar\psi^\p_{u^\p}(x^\p)=\psi_u(x)~~~u^\p=u
\end{eqnarray*}
Now we prove that,  $|U\rangle=|-u,-u^*\rangle$
in the following way. We have
$$
\frac{2s+1}{4\pi}\int{\d u}\langle n|U\rangle
\langle U|m\rangle~=~
\frac{2s+1}{4\pi}\int{\d u}
\langle m|u\,u^*\rangle\langle u\,u^*|n\rangle
{}~=~\delta_{nm}~.
\eqno(B.1)
$$
$$
\langle U|\gamma^\mu|U\rangle ~=~
\langle U|m\rangle\gamma^\mu_{mn}\langle n|U\rangle
{}~=~\langle u\,u^*|n\rangle\gamma^\mu_{mn}\langle m|
u\,u^*\rangle
{}~=~\langle u\,u^*|(\gamma)^T|u\,u^*\rangle~.
\eqno(B.2)
$$
We have, for integer spins $(\gamma^\mu)^T=-\gamma^\mu$ so that we get,
$\langle U|\gamma^\mu |U\rangle=-isu^\mu$
Also we have
$$
\langle U_1|U_2\rangle~=~\langle u_2u^*|u_1u^*\rangle
{}~=~e^{is\Omega[u_2,u_1,u^*]}\left(\frac{1+u_1\cdot u_2}{2}\right)^s~.
\eqno(B.3)$$
But we also we have
$$
\Omega[u_2,u_1,u^*]=
-\Omega[u_1,u_2,u^*]=
\Omega[-u_1,-u_2,-u^*]
$$
Also since $u_1\cdot u_2=(-u_1)\cdot(u_2)$ we can write:
$$
\langle U_1|U_2\rangle=e^{is\Omega[-u_1,u_2,-u^*]}
\left(\frac{1+(-u_1)\cdot(-u_2)}{2}\right)^{s}~.
$$
Eq.(B.1),(B.2) and (B.3) proves that,
$|U\rangle=|-u,-u^*\rangle ~.$

\noindent{\bf Case ii) $s$= half-odd integer spins:}

\noindent We have $x^\p$=x. Let
\begin{eqnarray*}
\psi^\p_{m^\p}(x^\p)&=&\bar\psi_{mm^\p}(x)C_{mm^\p}~,\nonumber\\
\bar\psi^\p_{m^\p}(x^\p)&=&C_{mm^\p}\psi_m(x)~,
\end{eqnarray*}
with $C^2=-1,~C^*=C=-C^T$. Then we have
\begin{eqnarray*}
\psi_u(x)&=&\langle uu^*|m\rangle\,\psi_m(x)\nonumber\\
&=&-\langle uu^*|m\rangle\,C_{mm^\p}\bar\psi_{m^\p}(x^\p)
\nonumber\\
&\equiv&\bar\psi_{m^\p}(x^\p)\langle m^\p|U\rangle~,
\end{eqnarray*}
where,
$$
\langle m^\p|U\rangle\equiv-\langle uu^*|m\rangle C_{mm^\p}~.
$$
Also we have,
\begin{eqnarray*}
\bar\psi_u(x)&=&\bar\psi_m(x)\langle m|uu^*\rangle\nonumber\\
&=&-\psi^\p_{m^\p}(x^\p)\,C_{mm^\p}\langle m|uu^*\rangle\nonumber\\
&=&-\langle U|m^\p\rangle\psi^\p_{m^\p}(x^\p)~.
\end{eqnarray*}
Here we have used the fact that,
$$
\langle U|m^\p\rangle=-C_{m^\p m}\langle m|uu^*\rangle~.
$$
Now again we will prove that,
$$
|U\rangle=|-u,-u^*\rangle~,
$$ through the following steps. First we have,
$$
%\label{aaa}
\frac{2s+1}{4\pi}\int\d u\langle m|U\rangle\langle U|n\rangle
{}~=~
\frac{2s+1}{4\pi}\int \d u(-)C_{nn^\p}\langle n^\p|uu^*\rangle\langle uu^*|
m^\p\rangle C_{m^\p m}
=\delta_{nm}\eqno(B.4)$$
%\label{bbb}
$$\langle U|\gamma^\mu|U\rangle
{}~=~-\langle uu^*|m^\p\rangle
C_{m^\p m}\gamma^\mu_{nm}C_{nn^\p}\langle n^\p|uu^*\rangle
{}~=~
-\langle uu^*|C(\gamma^\mu)C|uu^*\rangle
$$
$$\hskip4cm=~-is\,\gamma^\mu~,\eqno(B.5)$$
% \label{ccc}
$$\langle U_1|U_2\rangle
{}~=~-\langle u_2u^*|m^\p\rangle
C_{m^\p m}C_{mn}\langle n|u_1 u^*\rangle
 ~=~\langle u_2u^*|u_1u^*\rangle~.
\eqno(B.6)$$
Now from Eq.(B.4),(B.5) and (B.6) it follows that $
|U\rangle=|-u,-u^*\rangle~.$ So we have
\begin{eqnarray*}
\psi^\p_{u^\p}(x^\p)=\bar\psi_u(x)~~x^\p=x\nonumber\\
\bar\psi^\p_{u^\p}(x^\p)=\psi_u(x)~~~u^\p=-u
\end{eqnarray*}
Putting both results together, we have
\begin{eqnarray*}
\psi^\p_{u^\p}(x^\p)=(-1)^{2s}\bar\psi_u(x)~~~x^\p=x\nonumber\\
\bar\psi^\p_{u^\p}(x^\p)=\psi_u(x)~~~u^\p=-u
\end{eqnarray*}
\newpage
{\underline{4. Transformation of the propagator}}

We can now easily derive the transformation properties of the propogator
by looking at the bilinears. We sumarize the results in the next three
equations.\\
\noindent i) Proper Lorentz transformations\\
Under $x^\p=\Lambda x $ and $u^\p=u$ we have,
$$
\psi^\p_{u_1^\p}(x_1^\p)\bar
\psi^\p_{u_2^\p}(x_2^\p)=
\psi_{u_1}(x_1)\bar
\psi_{u_2}(x_2)~.
\eqno(B.7)
$$
ii) Parity\\
In this case $x^\p=-x$ and $u^\p=u$ and we have,
$$
\psi^\p_{u_1^\p}(x_1^\p)\bar
\psi^\p_{u_2^\p}(x_2^\p)=
-\psi_{u_1}(x_1)\bar
\psi_{u_2}(x_2)~.
\eqno(B.8)
$$
iii) Charge conjugation\\
Under charge conjugation we have, $x^\p=x$, $u^\p=-u$ and we get,
$$
\psi^\p_{u_1^\p}(x_1^\p)\bar
\psi^\p_{u_2^\p}(x_2^\p)=
\psi_{u_2}(x_2)\bar
\psi_{u_1}(x_1)~.
\eqno(B.9)
$$

\end{document}